\documentclass[twocolumn,floatfix,superscriptaddress,longbibliography,nobalancelastpage,aps,prb]{revtex4-2}
\usepackage[utf8]{inputenc}
\usepackage{amsmath,amssymb,latexsym,tikz}
\usepackage{amsfonts,amsthm}
\usepackage{mathptmx}
\usepackage{physics}    
\usepackage{graphicx}
\usepackage{ragged2e}
\usepackage{booktabs}
\usepackage{verbatim}
\usepackage[colorlinks,citecolor=blue,filecolor=black,linkcolor=blue,urlcolor=blue]{hyperref}    
\usepackage{cleveref}    
\Crefname{figure}{Fig.}{Figs.}
\Crefname{table}{Tab.}{Tabs.}
\Crefname{section}{Sec.}{Secs.}
\Crefname{equation}{Eq.}{Eqs.}
\definecolor{lime}{HTML}{A6CE39}
\DeclareRobustCommand{\orcidicon}
{
	\begin{tikzpicture} 
	\draw[lime, fill=lime] (0,0) circle [radius=0.15] node[white] {{\fontfamily{qag}\selectfont \tiny ID}};
	\draw[white, fill=white] (-0.0625,0.095) 	circle [radius=0.007];
	\end{tikzpicture}
	\hspace{-2.5mm}
}
\newcommand\orcidID[1]{\href{https://orcid.org/#1}{\orcidicon}}

\makeatletter
\renewcommand{\fnum@figure}{\textbf{FIG.~\thefigure}}
\renewcommand{\fnum@table}{\textbf{TABLE~\thetable}}
\makeatother

\begin{document}
		
	\title{Quantum metrology with partially accessible chaotic sensors}
	
	\author{Harshita Sharma\orcidID{0009-0007-6068-0115}}
	\email{harshita.research25@gmail.com}
	\affiliation{Department of Physics, Birla Institute of Technology and Science Pilani,
Pilani Campus, Vidya Vihar, Pilani, Rajasthan 333031, India.}
	\author{Sayan Choudhury\orcidID{0000-0001-7643-8111}}
	\email{sayanchoudhury@hri.res.in}
	\affiliation{Harish-Chandra Research Institute, a CI of Homi Bhabha National Institute, Chhatnag Road, Jhunsi, Allahabad 211019, India}
	\author{Jayendra N. Bandyopadhyay\orcidID{0000-0002-0825-9370}}
	\email{jnbandyo@gmail.com}
	\affiliation{Department of Physics, Birla Institute of Technology and Science Pilani,
Pilani Campus, Vidya Vihar, Pilani, Rajasthan 333031, India.}
	
\begin{abstract}
Most quantum metrology protocols harness highly entangled probe states and globally accessible measurements to surpass the standard quantum limit. However, it is challenging to satisfy these requirements in realistic many-body sensors. We demonstrate that both of these constraints can be overcome in quantum chaotic sensors. Crucially, we establish that even in the presence of partial measurement accessibility, chaotic dynamics enables initial unentangled states to exhibit Heisenberg scaling of the quantum Fisher information, $I_{\alpha}$ with time. In the weakly chaotic regime, we identify spin-coherent states placed at the edge of the regular islands in the mixed classical phase space as optimal initial states for enhanced sensitivity. On the other hand, in the strongly chaotic regime, $I_{\alpha}$ is insensitive to the choice of the initial state. Notably, quantum-enhanced sensitivity is achieved even when a very low fraction ($\sim 5\%$) of the qubits are accessible. These results establish quantum chaos as a robust resource for quantum-enhanced sensing under realistic accessibility constraints on accessibility.
\end{abstract}
	
\maketitle
	
\emph{\textbf{Introduction:}} In classical systems, the maximum achievable measurement precision for estimating an unknown parameter $\theta$ is governed by the Cram\'er-Rao bound~\cite{rao1945,Cramer1946}: $\Delta\theta\geq1/\sqrt{I_c}$, where $I_c$ is the Fisher information~\cite{Fisher1922,fisher1925}. We note that $I_c$ typically scales linearly with the number of classical resources $N$ and the duration of the protocol, $t$. These considerations lead to the standard quantum limit (SQL), $\Delta\theta\geq 1/\sqrt{Nt}$, which bounds the precision of classical measurements~\cite{giovannetti2006quantum,giovannetti2011advances}. Intriguingly, this limit can be surpassed by harnessing genuinely quantum resources, such as entanglement \cite{Giovannetti2004,Frowis2011,Boto2000,Dowling2008,Joo2011,Slussarenko2017}, spin squeezing \cite{kitagawa1993squeezed,Wineland1992,Ma2009,Hyllus2010,Gross2012,Rozema2014}, and coherence~\cite{lyu2020eternal,zou2025enhancing,biswas2025floquet,biswas2025discrete}. Consequently, quantum systems have emerged as natural candidates for metrological applications~\cite{montenegro2025quantum,agarwal2025quantum}, including gravitational-wave detection \cite{Goda2008,schnabel2010quantum,aasi2013enhanced,Danilishin2019}, particle physics experiments \cite{bass2024quantum}, navigation \cite{Giovannetti2001}, fundamental physics \cite{jackson2023probing,DeMille2024}, precision timekeeping~\cite{kaubruegger2025progress}, and magnetometry \cite{Baumgart2016,Mishra2022,Zhou2020}. In this framework, the ultimate achievable precision is governed by the quantum Fisher information (QFI), $I_{\alpha}$. The celebrated Heisenberg limit is achieved when the QFI scales quadratically both $N$ and $t$: $I_{\alpha} \propto N^2t^2$ \cite{toth2014quantum}. Several works have explored routes such as non-linearity~\cite{Luis2004,Beltran2005,Luis2007,Boixo2007,Boixo2008,Boixo2007,Beau2017,Yang2022,Roy2008,Pezze2009} and quantum criticality~\cite{zanardi2006,Invernizzi2008,Gammelmark2011,Skotiniotis2015,Rams2018,Chu2021,DiCandia2023,Yin2019,Yang2022,Zhang2022,Zanardi2008,PhysRevLett.121.020402,Hotter2024} to reach and even surpass this limit. However, the preparation and preservation of these states in many-body systems remains experimentally challenging. Furthermore, the optimal protocols for achieving the highest measurement precision often require globally accessible measurements, which are experimentally unfeasible. However, partial access to the system can lead to a drastic decrease in the QFI and, consequently, the sensing capabilities. These challenges motivate the need to devise mechanisms to enhance QFI that do not depend on the careful preparation of probe states or global accessibility. 
    
Intriguingly, some recent works have demonstrated that information scrambling in collective spin systems can be harnessed to circumvent the state preparation challenge~\cite{davis2016approaching,braun2018quantum,liu2021quantum,li2023improving,kobrin2024universal,hu2026quantum,montenegro2025enhanced}. Notably, quantum chaotic systems can be employed to achieve increased sensitivity even when the initial state is a product state~\cite{fiderer2018quantum,shi2025quantum}. Remarkably, these protocols remain robust in the presence of noise~\cite{schuff2020improving}, thereby providing a striking example of resilient designs~\cite{asthana2025projected}. Concurrently with these developments, some recent works have focused on the development of optimized measurement protocols that addresses the partial accessibility challenge in quantum critical and light-matter coupled systems~\cite{mishra2021driving,Adani2024,Wan2024,Montenegro2025,Li2024}. These considerations naturally raise an intriguing question: is it possible to harness chaos for {\emph {resilient}} quantum-enhanced sensing in the presence of partial access measurements? 
    
In this Letter, we present an affirmative answer to this question by demonstrating that long-range interacting chaotic systems provide a route for {\emph {resilient}} quantum-enhanced sensing even when a very small fraction of the system is accessible. We model the sensor as the paradigmatic quantum kicked top (QKT), where the strength of the chaoticity can be controlled by tuning the non-linearity~\cite{Haake1987}. This system serves as a powerful resource for quantum sensing, when globally accessible measurements can be performed~\cite{fiderer2018quantum}. Strikingly, we demonstrate that chaotic dynamics enables even the development of a large QFI, even when a a small fraction ($\sim 10\%$) of the system is accessible. We find that, for a moderately chaotic case, the QFI under partial accessibility exhibits a strong dependence on the choice of initial state. In particular, initial coherent states, localized at the edges of regular islands, exhibit a larger QFI at late times and achieve Heisenberg scaling in time. In the strongly chaotic regime, there is no initial state dependence and the temporal growth of $I_Q$ in the partially accessible QKT mirrors that of the fully accessible QKT. We conclude that quantum chaotic sensors retain their resilience in the presence of partial accessibility, thereby providing a robust platform for quantum metrology.\\

\begin{figure}
	 	\includegraphics[width=0.48\textwidth]{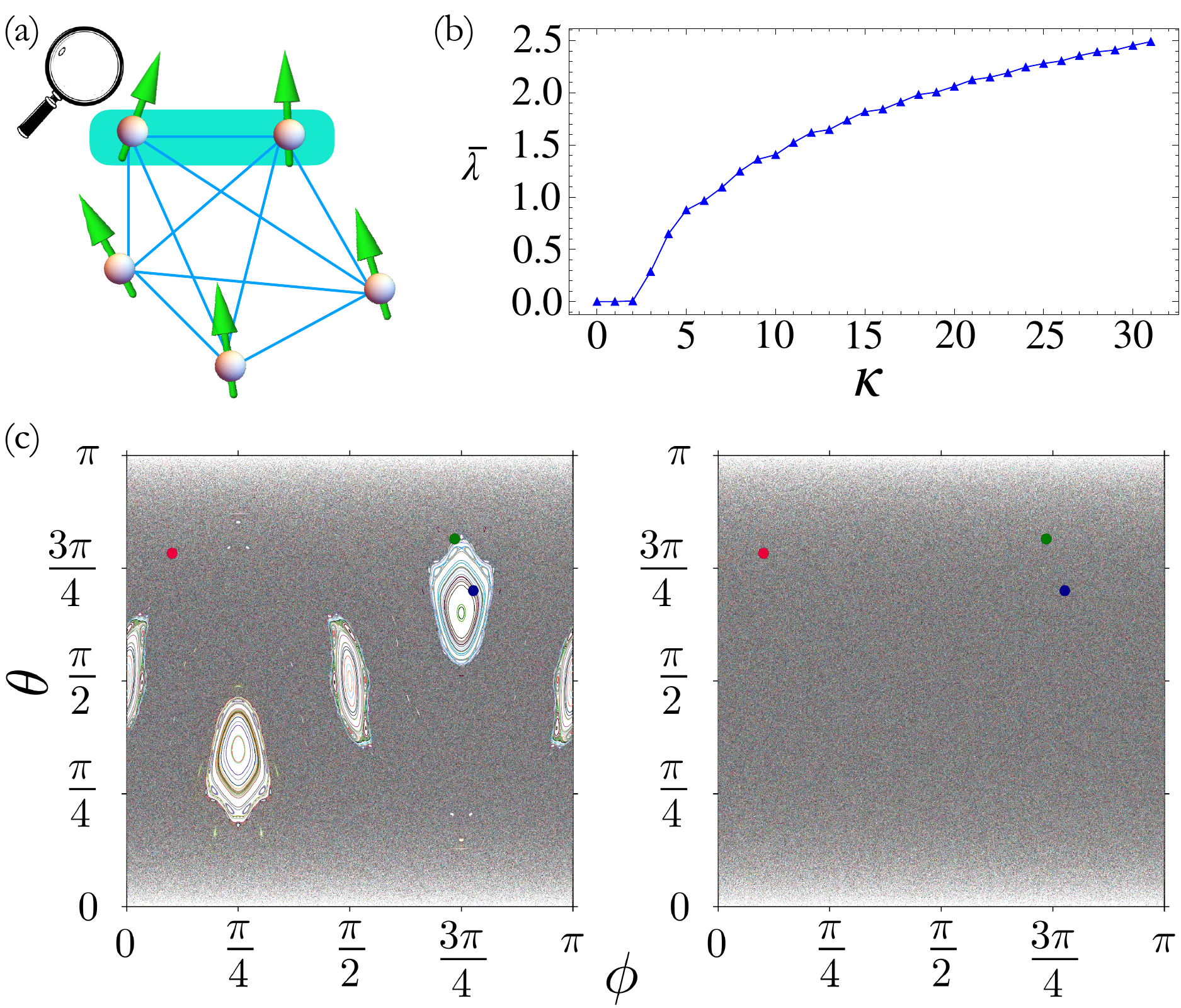}
	 	\caption{(a) A schematic illustration of the ``all-to-all" interacting quantum kicked top model under partial access. (b) The phase space averaged Lyapunov exponent that captures the sensitive dependence on initial conditions. The average is performed over 40,000 trajectories evolved over 50,000 drive periods. (c) The phase space map for (a) $\kappa = 3.0$ and (b) $\kappa = 30.0$. Different initial coherent states $\ket{\theta, \phi}$  are marked: non-equatorial island $\ket{2.20, 2.44}$ (blue circle), chaotic sea $\ket{2.46, 0.32}$ (red circle), and an edge state $\ket{2.56, 2.31}$ (green circle). Details regarding map generation are discussed in Appendix \ref{Map}. In the case of $\kappa = 30.0$, the system is strongly chaotic, and hence there are no distinguishable regular islands.}
	 	\label{Phasemap1}
	 \end{figure}

\emph{\textbf{Model and Diagnostics:}} We examine the dynamics of a periodically driven system of $N$ qubits interacting via infinite-range interactions described by:
	\begin{equation}
		H(t) = \frac{\kappa}{N}\sum_{i,j} s_i^y s_j^y + \frac{\alpha}{\tau} \sum_i s_i^z \sum_{n= -\infty}^{+\infty}{\delta(t-n\tau)},
		\label{eq1}
	\end{equation}
	where $s_{i}^{a} = \frac{1}{2} \sigma_{i}^{a}$, where $\sigma_{i}^{a}$ represent the standard Pauli matrices. the first term of the Hamiltonian describes the qubit-qubit interactions, and the second term describes a kick-induced rotation of each qubit about the $z$-axis. We set the kicking period $\tau=1.0$ and $\alpha = \frac{\pi}{2}$ for the rest of this work.  Due to the collective nature of the system, the Floquet unitary describing the stroboscopic time evolution of the system is represented by:
	\begin{equation}
		U = e^{-i\frac{\kappa}{2j}J_{y}^2}e^{-i\alpha J_{z}}
		\label{eq2}
	\end{equation}   
$J_i^{a} = \sum_{i} \sigma_i^{a}$ are the collective angular momentum operators satisfying $[J_i, J_k] = \iota\epsilon_{ikl}J_l$ and $j=N/2$. We note that time-evolution under this Floquet unitary has been studied widely in the context of the celebrated quantum chaotic model of the QKT~\cite{Haake1987, Piga2019}. Before proceeding further, it is instructive to examine the dynamics of the QKT in the (semi)classical limit ($j \rightarrow \infty$)  In this regime, the phase-space averaged Lyapunov exponent can be employed to characterize its chaoticity of the system. Our results are shown in Fig. \ref{Phasemap1} (b). It is evident that the system is not chaotic when $\kappa <2$; it begins to exhibit chaotic behavior as $\kappa$ increases. This is reflected in the phase space portraits of the system shown in Fig. \ref{Phasemap1} (c). In the moderately chaotic regime ($\kappa=3$), the system hosts a mixed phase space with co-existing regular islands and chaotic regions. However, in the strongly chaotic large $\kappa$ regime ($\kappa=30$), the regular islands completely disappear. In this work, we analyze the efficacy of the sensor in both of these regimes.
     
We characterize the sensing capabilities of the QKT sensor by computing the QFI~\cite{fiderer2018quantum}:
\begin{equation}
		I_\alpha(t) = 2 \sum_{nm}\frac{\vert{\bra{\psi_m} \partial_\theta \rho_\theta \ket{\psi_n}}\vert^2}{p_n + p_m},
		\label{eq6}
\end{equation} 
 where $\partial_\theta$ denotes differentiation with respect to the parameter $\theta$, $p_n$ and $p_m$ are eigenvalues of the the mixed state $\rho$ with eigenvectors $\ket{\psi_n}$ and $\ket{\psi_m}$, respectively, and the condition $p_n + p_m \neq 0$ is satisfied \cite{Braunstein1994,braunstein1996,Paris2009}. As noted earlier, $I_\alpha$ quantifies the precision with which a parameter, say $\theta$, encoded in a quantum state can be estimated by providing an upper bound on the Fisher information for all possible generalized measurements \cite{Braunstein1994}: $\Delta\theta_{CR}\geq \Delta\theta_{QCR} = 1/\sqrt{I_{q}}$~\cite{helstrom1967,Helstrom1969,Holevo1982}. We note that the QFI is also used as a diagnostic for multipartite entanglement \cite{Hyllus2012,Toth2012,Gietka2019} and for detecting quantum phase transitions~\cite{Yin2019,wang2014quantum,Poggi2024,Chakrabarti2025}. We note that for a pure state $\psi_\theta$, the general formula for the QFI simplifies to $I_\alpha(t) = 4\left( \bra{\partial_\theta \psi_\theta}\ket{\partial_\theta \psi_\theta} - \abs{\bra{\partial_\theta \psi_\theta} \ket{\psi_\theta}}^2 \right)$.
In the case of partial accessibility however, we compute $I_{\alpha}$ (Eq.~\ref{eq6}) for the reduced density matrix, $\rho_Q$ corresponding to the accessible $Q$ qubits of the $N-$qubit systems. The procedure for calculating $\rho_Q$ is explicitly provided in Appendix \ref{RDM}. In the rest of this work, we focus on the local estimation of $\alpha$.\\

    \begin{figure}
		\includegraphics[width = 0.49\textwidth]{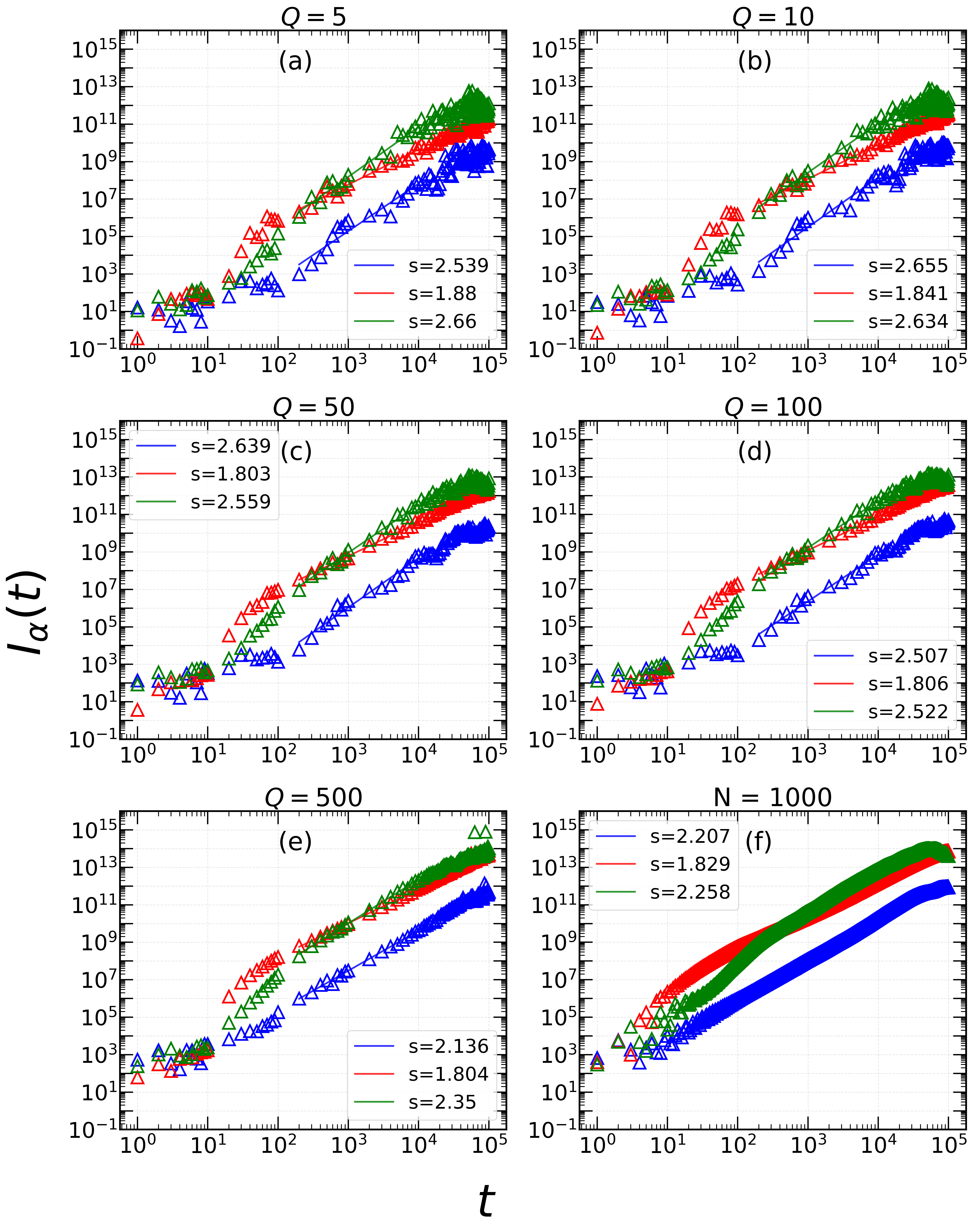}
		\caption{Scaling of the QFI, $I_\alpha$, is presented as a function of time $t$, for $\kappa = 3.0$ and different subsystem access $Q$, out of a total of $N =1000$ qubits. Each panel shows the QFI for three different initial coherent states: a non-equatorial island state $\ket{2.20, 2.44}$ (blue), a state at the chaotic sea $\ket{2.46, 0.32}$ (red), and an edge state $\ket{2.56, 2.31}$ (green). At sufficiently later times, the edge state exhibits the highest QFI, followed closely by the chaotic sea state. Importantly, even in only $\mathbf{5\%}$ access, the QFI for the edge state nearly reaches the same order of magnitude as the full access case. The QFI vs time plots for more initial states are shown in Appendix \ref{MR}.}
		\label{fig:QFI_t_k3}
	\end{figure}

    \begin{figure}
    	\includegraphics[height = 15cm, keepaspectratio]{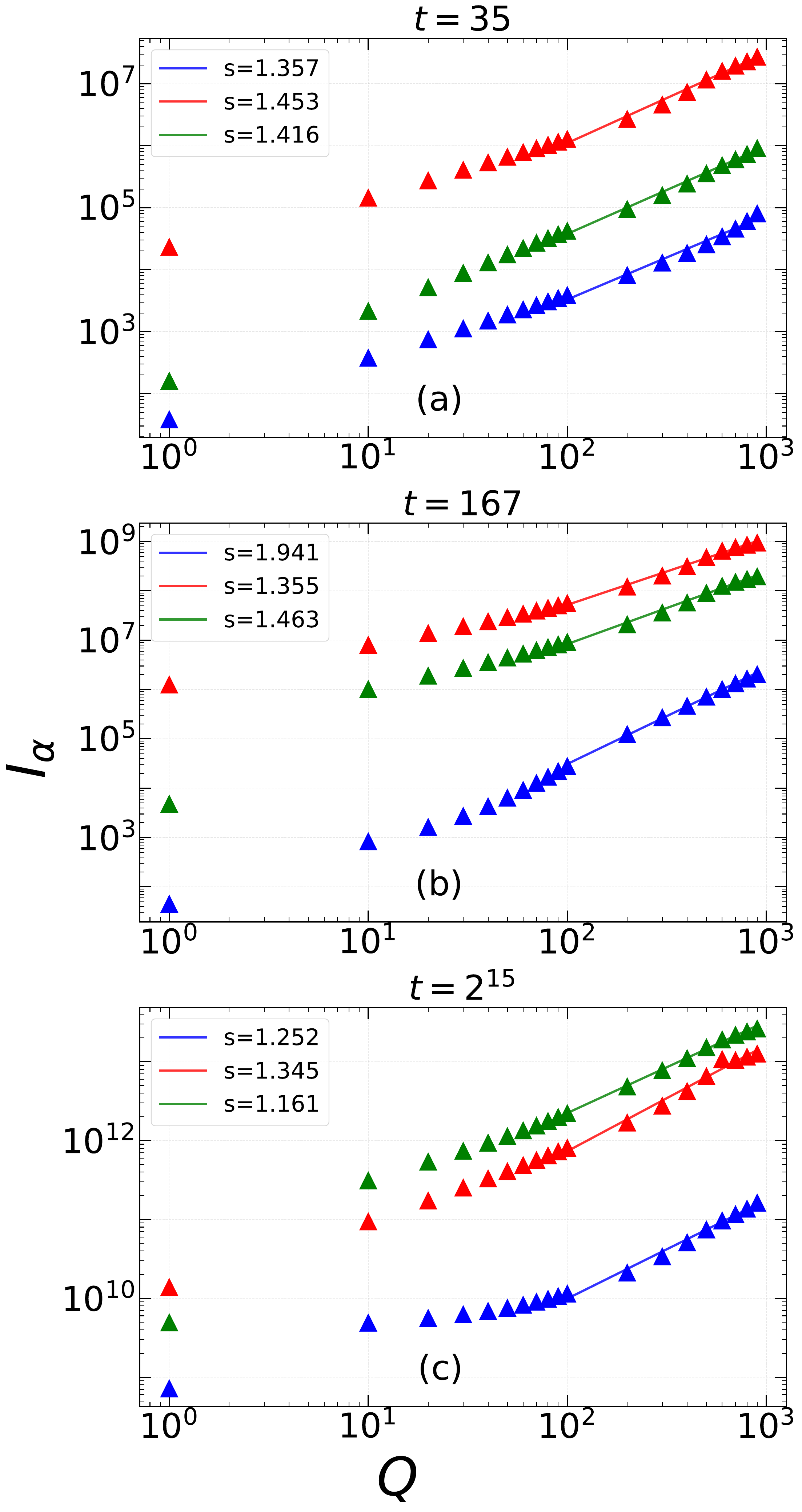}
    	\caption{Scaling of the QFI is shown as a function of the subsystem size $Q$ for $\kappa = 3.0$ and at three different times : (a) $t = t_E = 35$, the Ehrenfest time for the edge state. (b) $t = t_H = 167$, the Heisenberg time of the system. (c) $t = 2^{15}$, a much later time $t\gg t_H$. With time evolution,  the separation between the QFI of the initial chaotic and edge states reduces. Eventually, the QFI of the edge state surpasses that of the chaotic state and reaches its highest QFI value. A transition in QFI scaling is observed starting from $Q=100$. However, at short times, the chaotic initial states provide the larger QFI values and hence the best candidates for sensing. QFI vs $Q$ plots for more initial states are shown in Appendix \ref{MR}.}
    	\label{QKT_Q_k3_3T}
    \end{figure}
    
\emph{\textbf{Results:}} We now proceed to investigate the behavior of the QFI for the QKT sensor composed of $N = 1000$ qubits ($j=500$), both in the moderately chaotic ($\kappa = 3$) and the strongly chaotic ($\kappa = 30$) regime. We analyze the time-evolution of the system prepared in unentangled SU(2) spin-coherent states: \cite{Radcliffe1971,Arecchi1972,Haake2018}:
	 \begin{equation}
	 	\ket{\theta, \phi} = e^{i\theta\left(J_x \sin\phi - J_y\cos\phi\right)}\ket{j,j}
	 	\label{eq3}
	 \end{equation}

In the moderately chaotic regime, the QFI exhibits a strong initial state dependence~\cite{fiderer2018quantum}. To illustrate this, we have examined the dynamics of the system for 3 different initial states from the mixed phase space as shown in \Cref{Phasemap1}: a non-equatorial island state $\ket{2.20, 2.44}$, chaotic sea state $\ket{2.46, 0.32}$, and green for edge state $\ket{2.56, 2.31}$. These states are represented in blue, red, and green colors respectively, in all the figures.
	
The time-evolution of the QFI in this regime for different subsystem sizes $Q$, ranging from $Q = 5$ to $Q = 500$ is shown in \Cref{fig:QFI_t_k3} (a)-(h). The case of global accessibility is also shown for comparison in \Cref{fig:QFI_t_k3}(i). We note that at early times,  an initial state corresponding to the chaotic region exhibits the largest QFI value; however, at later times, it is overtaken by the edge state. For smaller values of $Q$, the QFI fluctuates significantly for different initial states. As the number of accessible qubits is increased however, these fluctuations are suppressed. Crucially, the dependence of the QFI on the initial states persists even when we increase the number of accessible qubits $Q$. This observation highlights the metrological advantage of chaotic and edge states over initial regular island states, even under partial access.  
    
For subsystem sizes up to $50\%$ accessibility, the QFI exhibits robust scaling with time, approximately resembling the fully accessible case. We note that the Ehrenfest time $t_E\simeq\ln(2J)/\lambda_{E}$ and the Heisenberg time $t_H \simeq J/3$ are characteristic time scales for the quantum dynamics of the QKT. We find that in the the time interval $t_H <t\leq 10^{4}$, the QFI becomes superextensive, and the scaling exponent lies in the range $s \simeq [1.803,2.719]$, indicating quantum-enhanced sensitivity. Moreover, the actual value of the QFI increases as $Q$ is increased, in accordance with expectations. However, for the edge state, the QFI nearly reaches the same order as the full accessible case with as little as $5\%$ access at later times. On the other hand, the overall growth of the QFI for all initial states with $50\%$ access closely follows the global access situation for $\kappa = 3.0$.
	
The dependence of QFI with different subsystem sizes $Q$ is shown in \Cref{QKT_Q_k3_3T} for three representative times, where $Q$ is varied from $1$ to $900$ qubits. At early times, the chaotic initial state yields the highest QFI for different subsystem sizes. At later times however, the difference between the QFI corresponding to the chaotic and edge states reduces, and eventually, the edge state overtakes the chaotic state. A noticeable change in the QFI scaling is observed from $Q = 100$ as seen in \Cref{QKT_Q_k3_3T}(b) and (c). These results suggest that, although the scaling of the QFI is sub-Heisenberg, it still exhibits a superextensive scaling for measurements involving at least a $10\%$ access to the total system.

\begin{figure}[t]
		\includegraphics[width=0.49\textwidth]{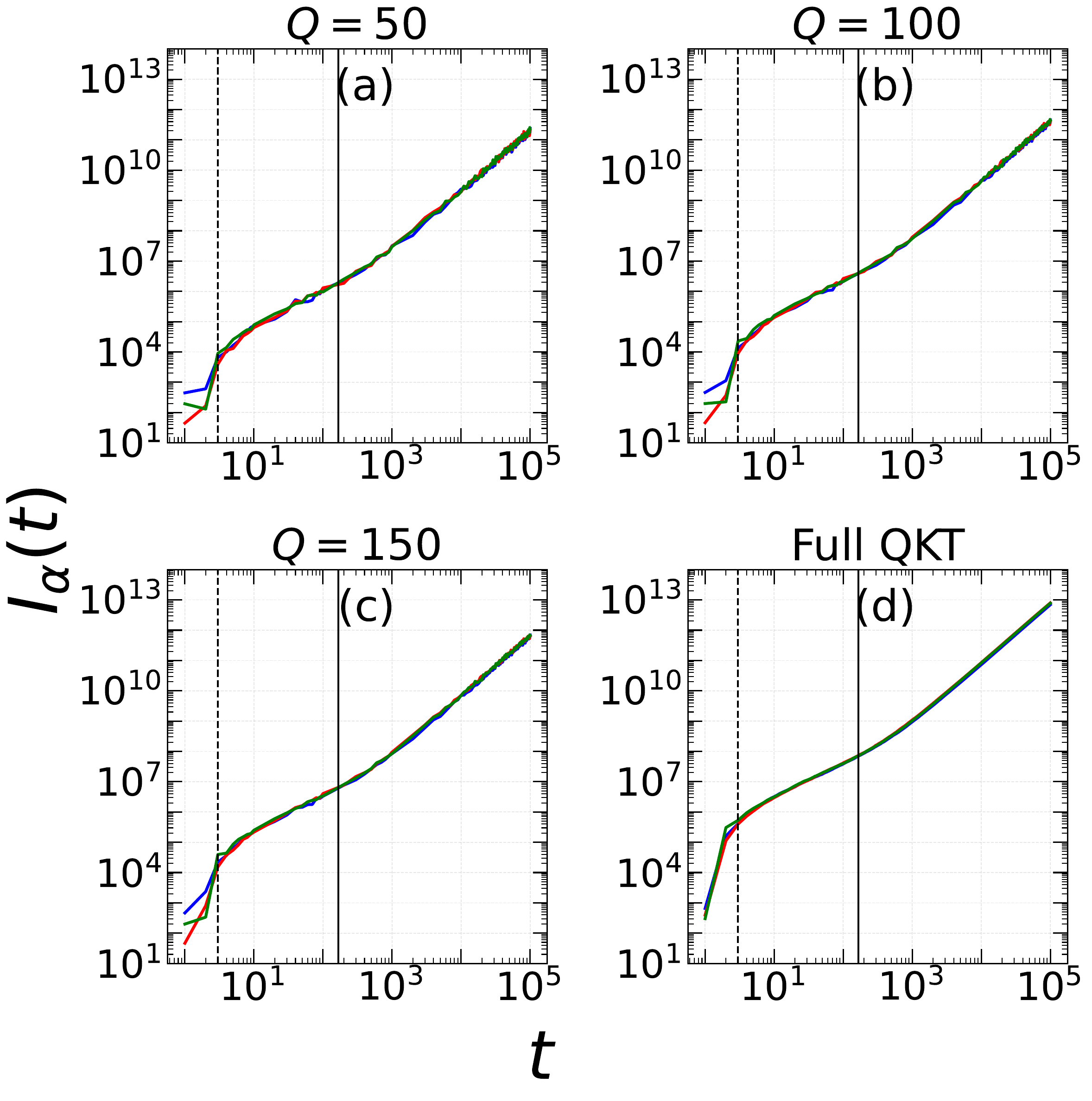}
		\caption{Scaling of the QFI as a function of time $t$ is shown at $\kappa = 30.0$ and for three different subsystem sizes: (a) $Q = 50$ qubits. (b) $Q = 100$ qubits. (c) $Q = 150$ qubits. (d) Fully accessed case with a total of $1000$ qubits. The dotted line represents the Ehrenfest time $t_E = 3$ and the solid line represents the Heisenberg time $t_H = 167$. The trends of the QFI with time for the partially accessible case behave the same as the fully accessible QKT for the large value of the chaoticity parameter $\kappa$.}
		\label{Fig3}
	\end{figure}

For strong chaos with $\kappa = 30.0$, the system rapidly loses the memory of its initial conditions, and the QFI does not exhibit any systematic initial state dependence~\cite{fiderer2018quantum}. This behavior persists even when only partial access to the system is available. Figure \ref{Fig3} provides evidence that the growth the QFI for different subsystem sizes $Q = 50,100,150$ is almost the same as the fully accessible case for strong chaos. In the partially accessible case, the QFI exhibits slight irregularity prior to the Ehrenfest time $t_E$, and after that, the QFI displays the same memoryless nature as the fully accessible case. Finally, we note that while the temporal scaling of the QFI differs across the three regimes: $t<t_E$, $t_E<t<t_H$, and $t>t_H$, the qualitative behavior of the QFI growth with the magnitude of QFI increasing systematically as more qubits are accessed. The QFI scaling is super-Heisenberg till $t = t_E$, superextensive for $t_E<t<t_H$ and near Heisenberg for $t>t_H$. 

\begin{figure}
		\includegraphics[width = 0.49\textwidth]{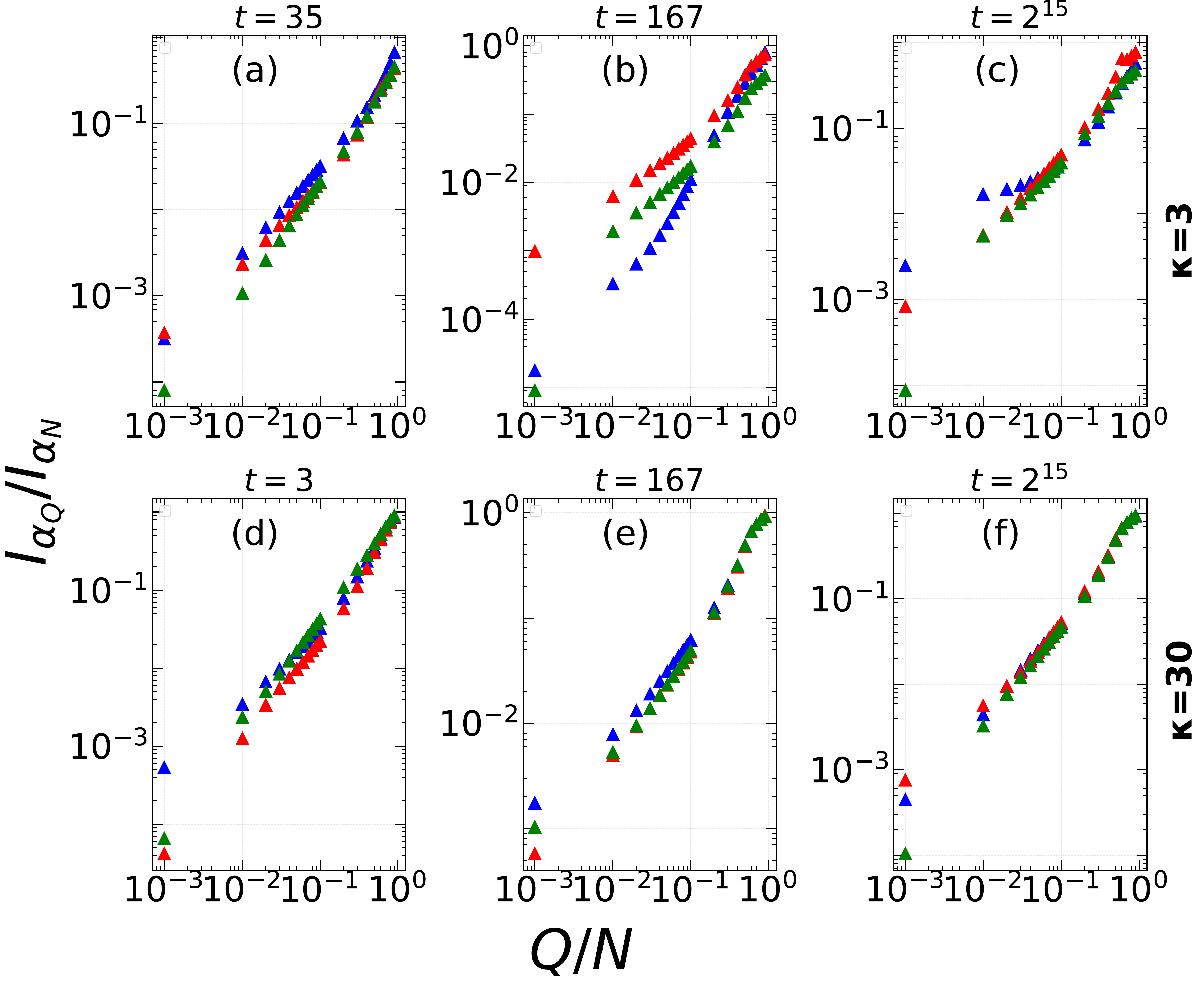}
		\caption{Fractional QFI versus fractional access $Q/N$ are shown for $\kappa = 3.0$ and $30.0$ at different times. Panels (a)-(c) are for $\kappa = 3.0$ case and (d)-(f) are for $\kappa = 30.0$ case. A transition in the slope is observed around $Q/N = 0.1$ for both the cases.}
		\label{Frac_qfi_j500}
\end{figure}

Finally, we characterize the sensing capabilities of the system for various amounts of partial accessibility. To this end, we examine the ratio of the QFI with partial access $I_{\alpha_Q}$, to that of the fully accessible case $I_{\alpha_N}$, for any given initial coherent state; we dub this measure `Fractional QFI'. \Cref{Frac_qfi_j500} shows the dependence of the fractional QFI on the fractional accessibility $Q/N$ at three characteristic times: $t=t_E, t_H$, and $t>>t_H$ for $\kappa = 3.0$ and $30.0$. A transition in slope is observed around $Q/N = 0.1$. The scaling for both $\kappa = 3.0$ and $\kappa = 30.0$ beyond this fractional access value (0.1) is super-extensive. Thus, access to about $10\%$ of the total qubits is sufficient to extract a substantial amount of information about the encoded parameter, particularly during early time evolution.  \\
    
\emph{\textbf{Summary and Outlook:}} In this Letter, we have demonstrated that long-range interacting chaotic spin systems serve as quantum-enhanced sensors, even under partial accessibility. Remarkably, even without global access, the QFI still exhibits Heisenberg scaling with time, showing that chaos-assisted metrological enhancement remains robust under realistic measurement constraints. This feature is particularly relevant for experimental implementations, where global access is often impractical and local measurements are employed instead. We note that the QKT has been realized in a variety of platforms including superconducting qubit processors~\cite{neill2016ergodic}, ultracold atomic systems~\cite{chaudhury2009quantum}, and NMR platforms~\cite{krithika2019nmr}, thereby making our predictions verifiable in near-term experiments.

There are several avenues for future work. A natural extension of this work would be to examine the resilience of these partially accessible chaotic sensors in the presence of decoherence. Furthermore, it would be intriguing to develop quantum control protocols to further enhance the sensing capabilities of this system. Finally, it would be interesting to extend these studies to other kinds of resilient quantum sensors, such as those based on projected ensembles~\cite{asthana2025projected}.\\
	
\emph{\textbf{Acknowledgments:}} H.S. and J.N.B. acknowledge the Department of Science and Technology (DST) for providing computational resources through the FIST program (Project No. SR/FST/PS-1/2017/30). SC acknowledges funding under the Government of India’s National Quantum Mission grant numbered DST/FFT/NQM/QSM/2024/3.
\clearpage
\onecolumngrid
\appendix

\section{Phase Map}
	\label{Map}
		
	The classical phase-space map can be obtained from the stroboscopic Heisenberg equations of motion for the generators, $X_i = \frac{J_i}{j}$ in the limit $j \to \infty$. For the Quantum Kicked Top (QKT) Hamiltonian, phase map is generated by the following equations:

   \begin{align*}
   	  X' &= (X\cos\alpha - Y\sin\alpha)\cos\!\big(\kappa(X\sin\alpha + Y\cos\alpha)\big)
	       + Z\sin\!\big(\kappa(X\sin\alpha + Y\cos\alpha)\big) \\
	Y' &= X\sin\alpha + Y\cos\alpha \\
	Z' &= -(X\cos\alpha - Y\sin\alpha)\sin\!\big(\kappa(X\sin\alpha + Y\cos\alpha)\big)
	+ Z\cos\!\big(\kappa(X\sin\alpha + Y\cos\alpha)\big)
   \end{align*}
	\begin{figure*}[b]
		\includegraphics[width=1\textwidth]{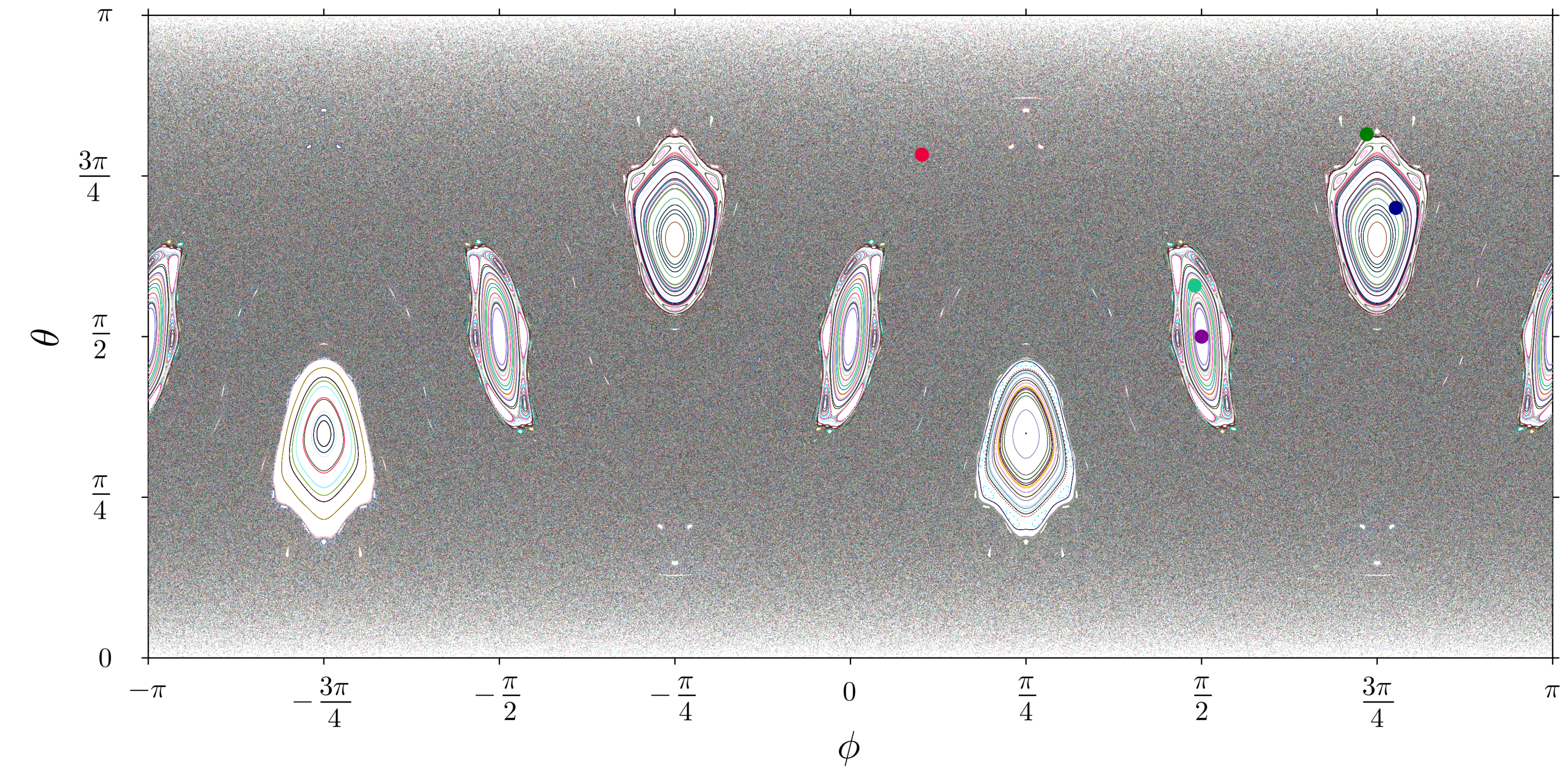}
		\caption{ Different initial coherent states $\ket{\theta, \phi}$  are marked: equatorial $\ket{\pi/2, \pi/2}$(purple circle) and $\ket{1.82, 1.54}$ (limegreen circle), non-equatorial $\ket{2.20, 2.44}$ (blue circle), chaotic sea $\ket{2.46, 0.32}$ (red circle), and an edge state $\ket{2.56, 2.31}$ (green circle). The color scheme of the different plots for these initial states will be maintained as mentioned here.}
		\label{FigS1}
	\end{figure*}   

	\section{Reduced Density Matrix}
	\label{RDM}
	The initial states used for simulation are the SU(2) spin-coherent states of the form :
     \begin{equation*}
		\ket{\psi}= \ket{j, \theta, \phi} = \ket{\theta, \phi} = e^{i\theta\left(J_x \sin(\phi) - J_y\cos(\phi)\right)}\ket{j,j}
	\end{equation*}
	
	These initial states can be expressed in the Dicke basis:
	\begin{equation}
	\ket{\psi} = \sum_{k = 0}^{N} c_{k} \ket{D_N^{(k)}}	
	\label{eqnS1}
    \end{equation}
   Now the Dicke states $\ket{D_N^{(k)}}$, which are completely symmetric $N-$qubit states of $k$ up-spins and $N-k$ down-spins, can be written in the bipartite form as \cite{Latorre2005,Nepomechie2024} :
   \begin{equation}
   	\ket{D_N^{(k)}}= \sum_{l = 0}^{Q} \sqrt{\frac{\binom{Q}{l}\,\binom{N-Q}{\,k-l\,}}{\binom{N}{k}}} \ket{D_Q^{(l)}} \otimes \ket{D_{N-Q}^{(k-l)}}
   	\label{eqnS2}
   \end{equation}
   Here we consider the bipartition into blocks of $Q$ and $N-Q$ spins. The $Q$ block has $l$ up-spins
   and block $N-Q$ has $k-l$ up-spins. From \Cref{eqnS1,eqnS2} we get:
   \begin{equation}
   	\begin{split}
   		\ket{\psi} &=\sum_{k = 0}^{N} \sum_{l = 0}^{Q} c_{k} \sqrt{\frac{\binom{Q}{l}\,\binom{N-Q}{\,k-l\,}}{\binom{N}{k}}} \ket{D_Q^{(l)}} \otimes \ket{D_{N-Q}^{(k-l)}}\\
   		&= \sum_{l = 0}^{Q} \sum_{r = 0}^{N-Q} c_{l+r} \sqrt{\frac{\binom{Q}{l}\,\binom{N-Q}{\,r\,}}{\binom{N}{l+r}}} \ket{D_Q^{(l)}} \otimes \ket{D_{N-Q}^{(r)}}\\
   	\end{split}
   	\end{equation}
	
	The $p,q^{th}$ element of the reduced density matrix, $\rho_{Q} = \trace_{N-Q}{\ket{\psi}\bra{\psi}}$ is given by:
	\begin{equation}
		(\rho_Q)_{p q} = \sqrt{\binom{Q}{p}\binom{Q}{q}}\sum_{r = 0}^{N-Q} \frac{\binom{N-Q}{r}}{\sqrt{\binom{N}{p+r}\,\ \binom{N}{q+r}}} c_{p+r} c_{q+r}^{*}
	\end{equation}

  \begin{figure*}[b]
	  	\includegraphics[width=0.8\textwidth]{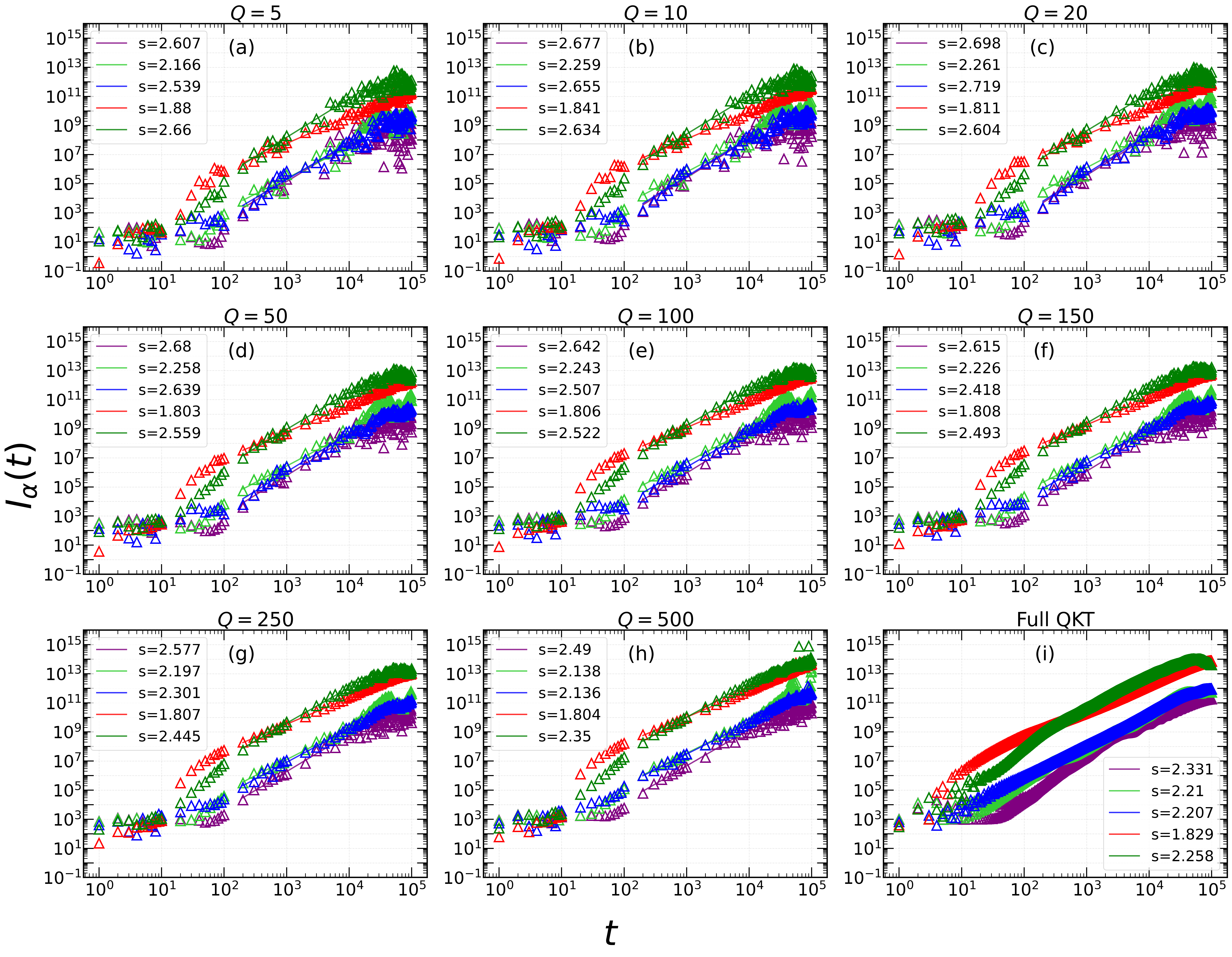}
	  	\caption{
        Scaling of the QFI $I_\alpha$ is presented as a function of time $t$, for $\kappa = 3.0$ and different subsystem access $Q$, out of a total of $N =1000$ qubits. At sufficiently later times, the edge state exhibits the highest QFI, followed closely by the chaotic sea state. Importantly, even in only $\mathbf{5\%}$ access, the QFI for the edge state nearly reaches the same order of magnitude as the full access case.}
	  	\label{FigS2}
	  \end{figure*}        
	  \section{More results}
	  \label{MR}
      In this section, we present the QFI results of more initial states to compare our results with a previous study \cite{fiderer2018quantum}. The color scheme for the QFI plots for the different initial points is fixed throughout the paper: purple for the equatorial island state $\ket{\pi/2, \pi/2}$, lime-green for island state $\ket{1.82, 1.54}$, blue for non-equatorial island state $\ket{2.20, 2.44}$, red for chaotic sea state $\ket{2.46, 0.32}$ and green for edge state $\ket{2.56, 2.31}$\\
 
The numerical results of the QFI for different $Q$ is
shown in \Cref{FigS2}. Among all the initial states, the chaotic state shows the largest QFI value initially but gets surpassed by the edge state at later times.
As seen the QFI for different subsystem upto $50\%$ accessibility shows a good scaling with time. Early time behaviour of the QFI is very random, there are fluctuations so the focus should be from the Heisenberg time, $t_H$. For time $t_H < t \leq 10^{4}$, the QFI shows quantum enhancement: on average, one can say the scaling is Heisenberg. While the edge state
QFI almost reaches the same order as the full top QFI in
as low as $5\%$ access at large times, the overall QFI trend
for all initial states in $50\%$ access almost mirrors the full
access case for $\kappa = 3$.
	  
	  \begin{figure*}[t]
	  	\includegraphics[width=0.8\textwidth]{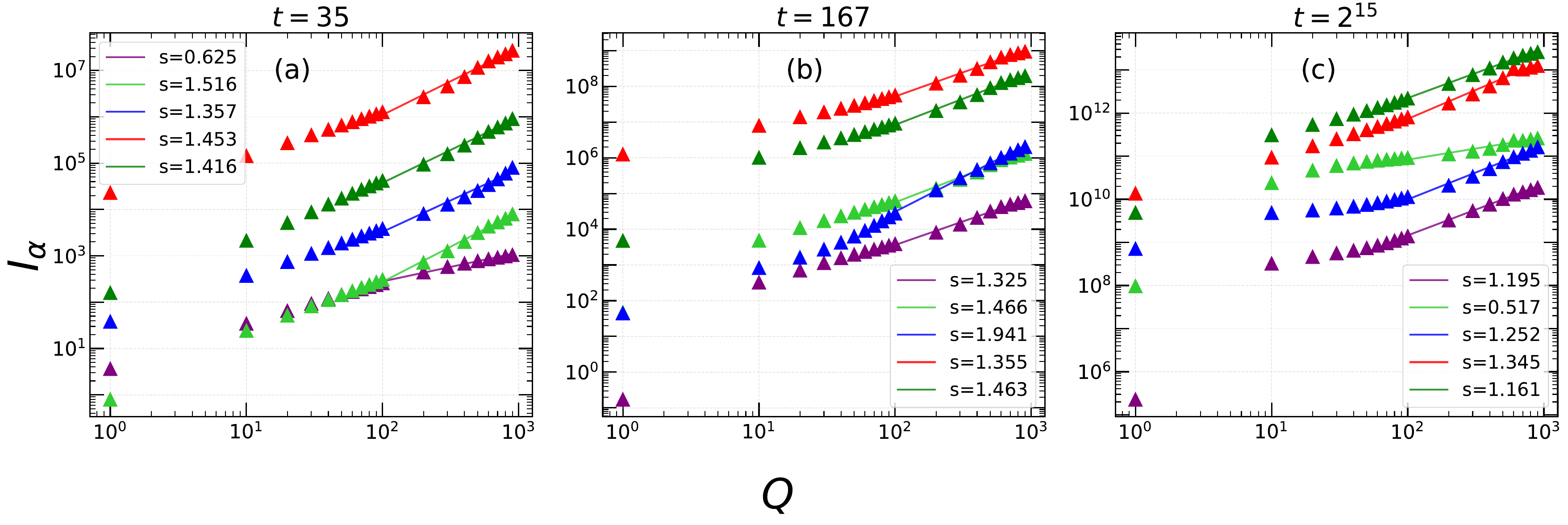}
	  	\caption{Scaling of the QFI is shown as a function of the subsystem size $Q$ for $\kappa = 3.0$ and at three different times : (a) $t = t_E = 35$, the Ehrenfest time for the edge state. (b) $t = t_H = 167$, the Heisenberg time of the system. (c) $t = 2^{15}$, a much later time $t\gg t_H$. With time evolution,  the separation between the QFI of the initial chaotic and edge states reduces. Eventually, the QFI of the edge state surpasses that of the chaotic state and reaches its highest QFI value. A transition in QFI scaling is observed starting from $Q=100$. However, at short times, the chaotic initial states provide the larger QFI values and hence the best candidates for sensing}
	  	\label{FigS3}
	  \end{figure*}
The QFI scaling with the subsystem size, $Q$ is shown in \Cref{FigS3} for three representative times. $Q$ is varied from 1 to 900 qubits here. The early time and late time behaviour of the different initial states are different. At early times, the chaotic initial state dominates in QFI value but the edge state overcomes it after the Heisenberg time. The QFI scaling with subsystem size, $I_{\alpha} \sim Q^{s}$ where $s\simeq [0.625,1.416]$ at the $t=t_E$. At $t=t_H$, $s$ lies in the range $\simeq [1.325,1.941]$. At very large time, $t=2^{15}$, $s\simeq [0.517,1.345]$. Overall, the QFI dependence on $Q$ can be considered to be linear. 
\twocolumngrid
\clearpage
\bibliography{ReferenceQKT}

\end{document}